# The DataFlow System of the ATLAS Trigger and DAQ

G. Lehmann, J. Bogaerts, M. Ciobotaru, E. Palencia Cortezon, B. DiGirolamo, R. Dobinson,
D. Francis, S. Gameiro, P. Golonka, B. Gorini, M. Gruwe, S. Haas, M. Joos, E. Knezo, T. Maeno,
L. Mapelli, B. Martin, R. McLaren, C. Meirosu, G. Mornacchi, I. Papadopoulos, J. Petersen,
P. de Matos Lopes Pinto, D. Prigent, R. Spiwoks, S. Stancu, L. Tremblet, P. Werner,
*CERN, Geneva, Switzerland*

M. Abolins, Y. Ermoline, R. Hauser
*Michigan State University, Department of Physics and Astronomy, East Lansing, Michigan*

A. Dos Anjos, M. Losada Maia
*Universidade Federal do Rio de Janeiro, COPPE/EE, Rio de Janeiro*

M. Barisonzi[1], H. Boterenbrood, P. Jansweijer, G. Kieft, J. Vermeulen
*NIKHEF, Amsterdam*

H.P. Beck, S. Gadomski[2], C. Haeberli, V. Perez Reale
*Laboratory for High Energy Physics, University of Bern, Switzerland*

M. Beretta, M.L. Ferrer, W. Liu
*Laboratori Nazionali di Frascati dell' I.N.FN., Fracasti*

R. Blair, J. Dawson, J. Schlereth
*Argonne National Laboratory, Argonne, Illinois*

D. Botterill, F. Wickens
*Rutherford Appleton Laboratory, Chilton, Didcot*

R. Cranfield, G. Crone
*Department of Physics and Astronomy, University College London, London*

B. Green, A. Misiejuk, J. Strong
*Department of Physics, Royal Holloway and Bedford New College, University of London, Egham*

Y. Hasegawa
*Department of Physics, Faculty of Science, Shinshu University, Matsumoto*

R. Hughes-Jones
*Department of Physics and Astronomy, University of Manchester, Manchester*

A. Kaczmarska, K. Korcyl, M. Zurek
*Henryk Niewodniczanski Institute of Nuclear Physics, Cracow*

A. Kugel, M. Müller, C. Hinkelbein, M. Yu,
*Lehrstuhl für Informatik V, Universität Mannheim, Mannheim*

A. Lankford, R. Mommsen,
*University of California, Irvine, California*

M. LeVine,
*Brookhaven National Laboratory (BNL), Upton, New York*

Y. Nagasaka,
*Hiroshima Institute of Technology, Hiroshima*

K. Nakayoshi, Y. Yasu,
*KEK, High Energy Accelerator Research Organisation, Tsukuba*

M. Shimojima,
*Department of Electrical Engineering, Nagasaki Institute of Applied Science, Nagasaki*

H. Zobernig,
*Department of Physics, University of Wisconsin, Madison, Wisconsin*

---

[1] Also Universiteit Twente, Enschede, Netherlands
[2] On leave from INP Cracow

**MOGT009**

This paper presents the design and prototype implementation of the DataFlow system of the ATLAS experiment. Its functional decomposition is described and performance measurements for each individual component are shown.

## 1. INTRODUCTION

The ATLAS experiment at the Large Hadron Collider (LHC) is scheduled to start taking data in 2007. Its main goals are the comprehension of the Electro-weak symmetry breaking mechanism and the discovery of new physics signatures beyond the ones predicted by the Standard Model [1]. The high event rate, due to the high luminosity of the collider, and the high cross-section for background processes as well as the large amount of data produced by ATLAS per event (~1.5 MB) requires the design of a performant Data AcQuisition (DAQ) system with three trigger levels. The first trigger level will carry out a rate reduction from 40 MHz down to at most 75 kHz. The second level trigger (LVL2) will reduce the rate by another two orders of magnitude, and the Event Filter (EF) will bring down the rate, at which data will be recorded, to of the order of 100 Hz.

The DataFlow system is responsible for moving data, which have passed the first level of selection to the High Level Triggers, and then for transferring the accepted data to mass storage. The High Level Triggers have been designed such that their requirements in terms of the bandwidth needed for data movement are similar. The second level trigger operates only on a fraction of the data (~2% of the full event), which has been tagged by the first level trigger as containing the relevant physics information (Regions Of Interest (ROI)). It has to be capable of handling events at up to 75 kHz and the average latency for the decision taking is of the order of 10 ms. The Event Filter on the other hand analyses the fully reconstructed events, but it operates at a rate of a few kHz (~ 2 kHz). Here the latency for decision taking is in the order of a few seconds.

The DataFlow is functionally decomposed in four building blocks: the ReadOut System (ROS), the ROI Collection, the Event Builder and the Event Filter I/O (EF I/O).

The ROS is responsible for receiving data from the detector, forward them on request to the second level trigger and Event Builder, and store the event data as long as it is explicitly told to delete them.

The ROI Collection is responsible for gathering the data required by the second level trigger.

The Event Builder is in charge of merging the event fragments coming from the ROS into a full event.

The EF I/O forwards events to the last selection stage, retrieves the accepted events form the Event Filter and puts them on mass storage.

## 2. THE FLOW OF DATA

When the first level trigger accepts and event, all front-end buffers push their data to the ROS via readout links. The standard protocol defined for the readout link is S-Link. It provides the transfer of 32 bit words at 40 MHz (160 MB/s), flow control and error detection with a bit rate lower than $10^{-12}$ [2]. ATLAS foresees to have 1628 of such links. The incoming bandwidth into the ROS will be of about 120 GB/s. The first level trigger sends the geometrical information related to the Regions Of Interest to the ROI Builder, which assembles the information from the calorimeter and muon triggers to form a single record, which will be used by the LVL2 for its analysis.

The ROI information is passed to one of the Level 2 Supervisors (L2SV): its task is to assign the event to the least loaded Level 2 Processing Unit (L2PU). The L2PU requests the data corresponding to the ROI from the ROS and checks whether the physical properties of the event satisfy any of the requirements set in the trigger menu.

The trigger decision is sent back to the L2SV, which forwards it to the DataFlow Manager (DFM) of the event building system. The summary information of the LVL2 trigger for accepted events is forwarded to a pseudo-ROS, which then participates in event building as part of the ROS. Accepted events are assigned to one of the Sub Farm Inputs (SFI); the SFI requires the full event from the ROS, assembles and formats it. On completion of the event building the SFI sends back a message to the DFM notifying that the event can be deleted. Event ids scheduled for deletion, the completed events as well as the LVL2 rejected ones, are grouped (typically a few hundreds) and then sent in a message to the ROS, to be eliminated from its buffers.

On demand from the Event Filter the SFI passes the full event to the last trigger selection stage. Here a complete reconstruction and offline like analysis are carried out. Accepted events are finally transferred to mass storage.

## 3. FUNCTIONAL DECOMPOSITION OF THE DATAFLOW

### 3.1. The ROS

The ROS receives data from the detector front end, on 1628 readout links. Its input bandwidth is in the order of 120 GB/s. All incoming fragments are stored individually in so called ReadOut Buffers (ROB).

The ROS provides the content of selected, individual ROBs to the LVL2 at a high rate (~2% of the ROBs at 75 kHz). Furthermore the ROS provides the data of all ROBs to the event building system at ~2 kHz.

The latency of the LVL2 is in the order of 10 ms while the one of the Event Builder is in the order of 100 ms: this requires each ROB to be capable of storing about 2.5 MB of data.

The ROS is split into $O(100)$ independent and identical modular units. Each is composed of three main building blocks: the ROBIn, the IOManager and the ROSController.

The ROBIn is a 64-bit PCI card interfacing to four read-out links and with four ROBs. The IOManager comprises all the software to control and drive the ROBIns as well as satisfy the requests coming from the High Level Triggers [3]. Depending on the final implementation of the ROBIn the data may either be sent directly to the requesting L2PU or SFI by the ROBIns via a dedicated network interface, or be collected first via the PCI bus and then sent by the IOManager (local event building). The ROSController encompasses all the functions which are not strictly related to the movement of data, such as interfacing to the ATLAS run control, accessing the configuration databases, connecting to the monitoring and error reporting system.

The performance of the present prototype of a ROS unit with 12 readout links is shown in figure a. Data are collected over the PCI and sent out over a single Gigabit Ethernet interface card. The fraction of events, which is accepted by the LVL2, is varied between 2 and 4 %. For a large fraction of the ROBs being requested by the LVL2 it is possible to see that the ROS becomes almost insensitive to the amount of event building. Here the performance is completely dominated by the rate at which the IOManager can handle the incoming ROI requests (asking for data of individual ROBs) over the network, for the 12 ROBs.

It is possible to see that already with today's technology the performance requested by ATLAS is achievable.

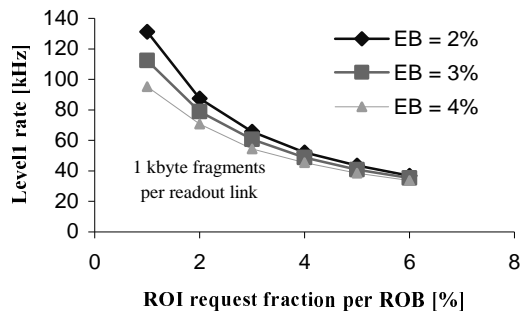

Figure a: First level trigger rate sustained by a ROS unit containing 12 readout links as a function of the amount of data requested for LVL2 processing. A constant percentage of events undergoes event building. The data from the 12 links is collected via the PCI bus and then sent out on a single gigabit Ethernet network interface card using the UDP protocol. These measurements were carried out on a 2 GHz single processor PC.

### 3.2. ROI Collection

The ROI Collection is the part of the DataFlow which provides data and ROI information to the LVL2. It receives the ROI information from the first level trigger at 75 kHz, forms a ROI record per event, retrieves ROI data from the ROS (~2% of the full event) and forwards the LVL2 decision to the Event Builder. It has been factorized in five components: the ROI Builder, the L2SV, the L2PU, the pseudo-ROS and the Local Controller.

The ROI Builder is a custom build 9U VME module with a flexible number of S-Link inputs and outputs; all other components are software applications running on conventional PCs [3]. The number of output links from the ROIBuilder is determined by the number of L2SVs needed to handle an event rate of 75 kHz. At present, measurements on a prototype implementation of the L2SV (based on a dual Pentium Processor clocked at 2.4 GHz) show that one supervisor can handle up to 30 kHz, so that there need to be at least three L2SV to handle the ATLAS rate requirements.

The pseudo-ROS is a non-demanding application which will have to receive and forward a few kB of information at event building rate (~ 2 kHz). The most critical application of the ROI Collection subsystem is the L2PU, which has to deal both with I/O and with the execution of analysis algorithms.

Figure b shows the I/O performance of a L2PU as a function of the size of the ROI. It is visible that the time requested for I/O is very small compared to the estimated 10 ms of trigger latency. The plot also shows how the time to perform the ROI collection changes when varying the number of ROS units over which a single ROI is distributed. The curves indicate that the mapping of the readout links onto the individual ROS units has a large impact on the time requested for the ROI Collection and should be optimized in order to minimize the distribution of the ROIs over ROS units.

From a point of view of pure data flow less than 100 L2PUs would already be sufficient for ATLAS.

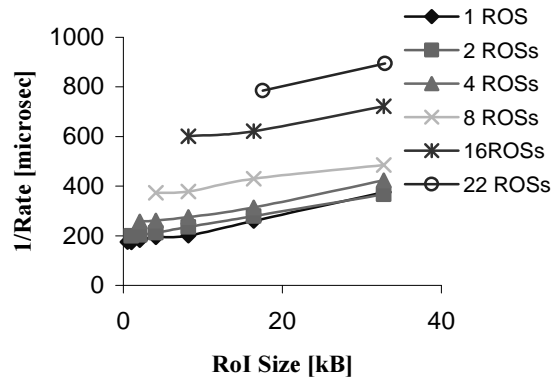

Figure b: Time requested by a L2PU application to collect the ROI information as a function of the data size. Several curves are shown, depending on how many ROS units have to addressed to gather the information.

### 3.3. The Event Builder

The Event Builder gathers all ROB fragments of LVL2 accepted events and builds complete, formatted events. It has been decomposed in three software applications [3] running on a variable number of PCs: the DFM, the SFI, and the Local Controller. The number of PCs requested for event building is determined by the rate at which each SFI can assemble events. The Local Controller interfaces the Event Builder to the ATLAS run control, while each application directly connects to the error handling and monitoring system, as well as it accesses the configuration database.

The task of the DFM is to provide load balancing between the active SFIs, thus insuring an effective use of the resources. It forwards the event id of an accepted event to a chosen SFI and waits for the completion of the event building process. The SFI gathers the fragments residing in the ROS, using a request/response protocol, which optimizes the traffic in the event building network [4]. A system of timeouts, in the DFM as well as in the SFI, prevents the event builder to stop running in case of lost event fragments or control messages.

Figure c shows the rate sustained by a single SFI as a function of the amount of local event building provided by an individual ROS unit. The SFI is able to exploit a bandwidth of 95 MB/s for event building. These measurements were carried out on a dual Pentium processor, clocked at 2.4 GHz.

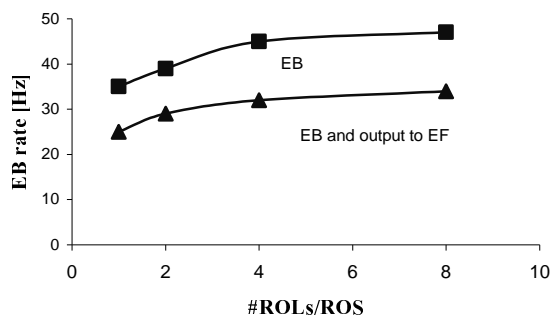

Figure c: Event building rate of an individual SFI as a function of the grouping of links done in a ROS unit. The rate increases when grouping the readout links, because the SFI has to send less request messages and receives back larger data fragments.

### 3.4. The EF I/O

The EF I/O is in charge of providing complete events to the Event Filter and to put accepted events to mass storage. It has been decomposed into an EF I/O library which implements the communication protocol between DataFlow and Event Filter and one software application, the Sub Farm Output (SFO), which is in charge of recording the data. The SFO is a non-demanding application since only ~10% of the events sent by the SFI to the trigger will have to be retrieved and stored. On the other hand the EF I/O library itself is critical since it is used by every SFI and shares the same CPU resources of the Event Builder. Figure c shows the degradation in Event Builder performance when the EF I/O is switched on. The highest event rate sustained by each SFI drops to about 35 Hz, which means that at least 60 SFIs will be needed to satisfy ATLAS requirements. This number is not only justified by the present measurements but also by the fact that for safety reasons a single link of the event building network will not be exploited at more than 60-70% of its nominal bandwidth.

### 3.5. Conclusions and Outlook

We have shown that the present prototype implementation of the ATLAS DataFlow system is locally capable of sustaining all the requirements set by the experiment. Now all components are sufficiently mature it will be possible to operate the prototype DataFlow system as a whole, optimize it towards performance and study its scaling issues.